\documentclass[pre,final,twocolumn,showpacs,showkeys]{revtex4}
\usepackage{amsmath}
\usepackage{graphicx}
\usepackage{amsfonts}
\usepackage{amssymb}
\usepackage{subfigure}

\def\identity{\leavevmode\hbox{\small1\kern-3.2pt\normalsize1}}

\begin{document}

\title{Equipartitions and a Distribution for Numbers: A Statistical Model for Benford's Law}

\author{Joseph R. Iafrate$^{1,2}$}
\author{Steven J. Miller$^2$}
\author{Frederick W. Strauch$^1$} \email[Electronic address: ]{Frederick.W.Strauch@williams.edu}
\affiliation{$^1$Department of Physics, \\
$^2$Department of Mathematics and Statistics, \\ Williams College, Williamstown, MA 01267}

\date{\today}

\begin{abstract}
A statistical model for the fragmentation of a conserved quantity is analyzed, using the principle of maximum entropy and the theory of partitions.  Upper and lower bounds for the restricted partitioning problem are derived and applied to the distribution of fragments.  The resulting power law directly leads to Benford's law for the first digits of the parts.
\end{abstract} 
\maketitle

\newcommand{\Q}{\mathbb{Q}}
\newcommand{\N}{\mathbb{N}}
\newcommand{\Z}{\mathbb{Z}}
\newcommand{\E}{\mathbb{E}}
\newcommand{\Ord}{\mathcal{O}}
\newcommand{\bb}{\begin{eqnarray}}
\newcommand{\ee}{\end{eqnarray}}
\newcommand{\var}[1]{{\rm Var}\left[#1\right]}

\newtheorem{thm}{Theorem}[section]
\newtheorem{lemma}[thm]{Lemma}

\newcommand{\subs}[1]
{ 
	\mbox{\scriptsize{#1}}
}

\section{Inroduction}

Given an arbitrary collection of numbers, can one predict what their distribution will be?  It would be fanciful to think so, but consider the following rationale.  Any experiment in nature involves partitioning some part of the universe (in mass, energy, volume, or some other physical quantity) from the rest.  The sizes of the resulting numbers are then subject to the conservation laws of physics (mass-energy, momentum, angular momentum).  Indeed, in statistical physics there is a fundamental distribution of energy $E$, the celebrated Boltzmann distribution $e^{-E/kT}$, where $T$ is the temperature and $k$ is Boltzmann's constant.  This distribution, while fully justified only in certain thermodynamic limits \cite{PathriaText}, is an extraordinarily powerful tool for analyzing many systems.  Could there be such a fundamental distribution of numbers?

Surprisingly,  many distributions in nature, economics, and sociology, such as Zipf's law, follow power-law distributions rather than the exponential Boltzmann distribution.  Such power laws have inspired a wide variety of explanations and arguments over the years \cite{Newman05}.  Most recently, arguments based on information theory known as random group \cite{Baek11} or community \cite{Peterson13} formation have shown how long-tailed distributions with general power laws can be derived from a small parameter model.  These are written in terms of a maximum entropy principle \cite{Visser13} given simple constraints.  In physics, however, such a procedure \cite{Jaynes57}, using the constraint of energy conservation, leads to the Boltzmann distribution.  How then could a physical conservation law lead to a power-law distribution?  

In this paper we argue that, under certain conditions similar to energy conservation, there is indeed a universal distribution that is intimately related to the Boltzmann distribution for quantum particles.   In an appropriate limit we call the {\em equipartition limit}, this distribution tends to the simplest inverse power law.   We further provide a concrete combinatorial proof of this limit, and verify it against numerical simulations.  Most importantly, we show how this limit has, as a simple consequence, Benford's law for the leading digit distribution \cite{Raimi76}.   A data set is Benford if the probability of observing a first digit of $d$ in {1, 2, ..., 9} is $\log_{10}(1 + 1/d)$.  This property has been observed in a wide variety of data sets from economics, sociology, mathematics, physics, geology, among others \cite{MillerBook}.  While many mathematical processes are known to exhibit the Benford property \cite{Berger11}, we believe that this argument, originally due to Lemons \cite{Lemons86}, is one of the simplest.  In short, a conservation law implies a power law that directly leads to Benford's law.  

This paper is organized as follows.  We begin in Section II by showing how the principle of maximum entropy, when applied to the partition of numbers, leads to a power law for the average number of parts of a given size.  This is extended in Section III, in which the theory of partitions is used to justify this result, in an appropriate limit.   The implication of this power law for Benford's law is presented in Section IV, while an extension to more general power laws is presented in Section V.  We conclude in Section VI, and provide additional mathematical details in the Appendix.

\section{Power Law from Maximum Entropy}

The main topic of this paper is the distribution of parts subject to an overall conservation law.  In more detail, we consider the distribution of $N$ numbers $n_j$, corresponding to piece sizes $x_j$, so that the total pieces add up to some given quantity $X$:
\begin{equation}
X = \sum_{j=1}^N n_j x_j.
\label{conv_law}
\end{equation}
Here the part set $\{x_1, x_2, \dots, x_N\}$ is fixed, but the number $n_j$ of parts of a given size $x_j$ is not.  The numbers $n_j$ specify a partition of $X$, which could result from a fragmentation process, as might occur in nuclear physics \cite{Sobotka85}.  We consider the set of of all such partitions of a quantity $X$, subject to Eq. (\ref{conv_law}).  The distribution we seek corresponds to the average number of parts $\langle n_j \rangle$ of size $x_j$, when all partitions can occur with equal probability.  The essence of this argument was originally given by Lemons \cite{Lemons86}, using heuristic arguments for a continuous set of parts.  In this section we will derive the probability distribution and average number for a discrete part set by using methods from statistical physics, namely Jaynes's principle of maximum entropy \cite{Jaynes57}.  A similar approach, specific to the fragmentation of solids, can be found in \cite{Englman91}, while an alternative application of maximum entropy to Benford's law can be found in \cite{Kafri2009}.  

In this formulation, we look for the probability distribution $p(\vec{n})$ for finding $n_1$ pieces of size $x_1$, $n_2$ pieces of size $x_2$, etc., that maximizes the entropy
\begin{eqnarray}
\mathcal{S} = - \sum_{\vec{n}} p(\vec{n}) \ln p(\vec{n})  - \alpha \left( \sum_{\vec{n}} p(\vec{n})- 1\right) \nonumber \\
 - \beta \left( \sum_{\vec{n}}  p(\vec{n}) \left[\sum_j n_j x_j - X \right]\right),
\end{eqnarray}  
where $\vec{n}$ is the vector of integers $(n_1, n_2, \dots, n_N)$.  Maximizing the entropy, we find
\begin{equation}
p(\vec{n}) = \prod_{j=1}^N (1 - e^{-\beta x_j}) e^{-\beta n_j x_j},
\end{equation}
where $\beta$ is a Lagrange multiplier, to be specified below.  This result conforms with the usual expectation that the distribution associated with a conserved quantity is exponential. 

Given the probability distribution for $\vec{n}$, we now consider how frequently each part $x_j$ occurs.  That is, when we observe a given partitioning of a system, we find a number of parts $n_1$ of size $x_1$, $n_2$ of size $x_2$, $n_3$ of size $x_3$, etc.  The distribution of observed part sizes (or fragments) $x_j$ will be proportional to the number of occurrences of a given part $x_j$, namely $n_j$.   Thus, we compute the expectation value of $n_j$ as a function of $x_j$:
\begin{equation}
\langle n_j \rangle = \sum_{\vec{n}} n_j p(\vec{n}) = \frac{1}{e^{\beta x_j} - 1},
\label{navg1}
\end{equation}
where the Lagrange multiplier is found by the conservation equation
\begin{equation}
\sum_j \langle n_j\rangle x_j = \sum_j \frac{x_j}{e^{\beta x_j} - 1} =  X.
\label{navg2}
\end{equation}
This form of the fragment distribution is formally equivalent to the average number of quanta for a set of harmonic oscillators with energies $x_j = \hbar \omega_j$ and inverse temperature $\beta = 1/kT$ \cite{PathriaText}.  

In the limit when $X \gg 1$, as is usually the case in real-world data, we expect that $\beta \ll 1$; this corresponds to a high-temperature limit.  In this case we can solve the conservation equation perturbatively in $\beta$ to find $\beta \approx N/X$ and thus
\begin{equation}
\langle n_j \rangle \approx \frac{1}{\beta x_j} \approx \frac{X}{N x_j}.
\label{keyeq1}
\end{equation}
We call this the {\em equipartition limit}, by analogy with the high-temperature limit for quantum harmonic oscillators, in which each oscillator has the same average energy $\langle n_j\rangle \hbar \omega_j = k T$.  

\begin{figure}
\includegraphics[width=3 in]{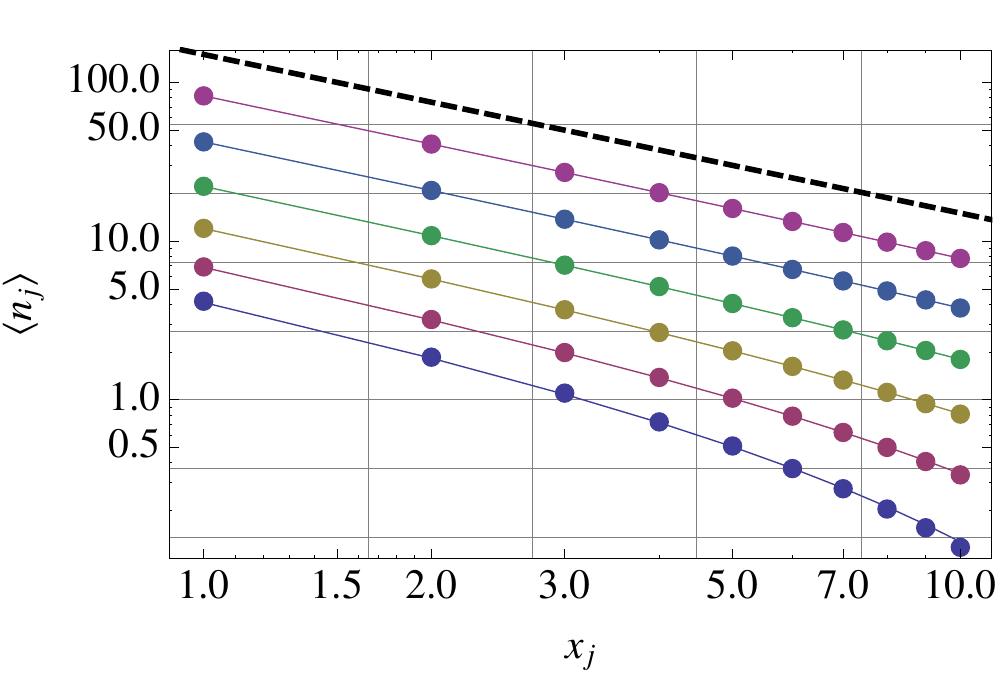}
\label{onlyfig}
\caption{Average number of parts $\langle n_j\rangle$ for $\{ x_j \} = \{1, 2, 3, 4, 5, 6, 7, 8, 9, 10\}$, for various values of $X = \{25, 50, 100, 200, 400, 800\}$ (from bottom to top).  The dots are exact calculations and the solid lines the maximum entropy result using Eq. (\ref{navg1}).  Also shown is the equipartition result of Eq. (\ref{keyeq1}), for $X = 1600$ (dashed).}
\end{figure}

This derivation provides solid evidence for Lemons' original argument \cite{Lemons86} that the number of parts of size $x$, when subject to a conservation law of the form Eq. (\ref{conv_law}), satisfies the power law $n(x) \sim 1/x$.  The approach given here also applies to more general partitions, including those with a continuous set of parts.   It can also be tested numerically.  Using the part set $\{1, 2, \dots, 10\}$, we have calculated the exact set of partitions for values of $X$ between 25 and 800 (the latter with over 100 trillion partitions) and the corresponding average values for $\langle n_j\rangle$.  The results, shown in Fig. 1, are well described by the maximum entropy result Eq. (\ref{navg1}), provided we numerically solve Eq. (\ref{navg2}) for $\beta$.  Finally, these results converge to the equipartition result Eq. (\ref{keyeq1}) for large $X$.

\section{Partition Number Calculation}
In the previous section we presented what could be termed a ``canonical ensemble'' calculation of the fragment distribution $n(x)$.  Such a calculation applies to the behavior of a set of systems for which the conservation law holds on average.  An alternative calculation uses the ``microcanonical ensemble'', a set of systems for which the conservation law holds exactly \cite{PathriaText}.   In this section we consider such a calculation of the equipartition limit of Eq. (\ref{keyeq1}), using the theory of integer partitions \cite{AndrewsText}.   In this framework we can find the average number of parts by exactly averaging over all partitions of the conserved quantity.

We note that {{\em unrestricted }} partition problems have been used previously to analyze fragmentation of nuclei \cite{Sobotka85, Mekjian90, Lee92,Chase94,Botvina2000}.  By contrast, here we consider the {{\em restricted}} partition number $P_H(X)$, that is, the number of ways to partition an integer $X$ into the set of integers $H = \{ x_1=1, x_2, \dots, x_N\}$, as in Eq. (\ref{conv_law}).   Note that setting $x_1 = 1$ ensures that a partition will exist for every $X$.

We begin by introducing the generating function
\begin{equation}
\label{eq:genfun1}
\sum_{k = 0}^{\infty} P_{H}(k) q^{k} = \prod_{x \in H} (1 - q^{x})^{-1},
\end{equation}
from which any given number can be obtained by multiple differentiation:
\begin{equation}
\label{eq:totalparts}
P_{H}(X) = \frac{1}{X!} \times \left( \frac{\partial}{\partial q} \right)^{X} \prod_{x \in H} (1 - q^{x})^{-1} \bigg|_{q = 0}.
\end{equation}
Evaluating this partition number is a hard problem, but useful approximations  \cite{Nathanson2000,Almkvist2002} and bounds \cite{Colman87,Agnarsson2002} exist.  

The average number of parts of size $x_j$ can be found by manipulating the partition functions $P_H(X)$ and $P_H(X;n_j)$:
\begin{equation}
\langle n_j \rangle = \frac{1}{P_H(X)} \sum_{n_j} n_j P_H(X; n_j),
\end{equation}
where $P_H(X;n_j)$ is the number of partitions of $X$ with exactly $n_j$ parts of size $x_j$.  Using its generating function, 
\begin{equation}
\sum_{k=0}^{\infty} P_H(k; n_j) q^k = q^{n_j x_j} \prod_{x \ne x_j} (1-q^x)^{-1}
\end{equation}
and performing the sum over $n_j$, we have
\begin{equation}
\sum_{n_j} n_j P_H(X; n_j ) = \frac{1}{X!}  \times \left( \frac{\partial}{\partial q} \right)^{X} \frac{q^{x_j}}{1-q^{x_j}} \prod_{x \in H} (1 - q^{x})^{-1} \bigg|_{q = 0}.
\end{equation}
This last expression can be simplified by using the Leibniz Rule for differentiation, so that
\begin{equation}
\langle n_j \rangle = \frac{1}{P_{H}(X)} \sum_{\ell = 1}^{\lfloor X/x_j \rfloor} P_{H}(X - \ell x_j).
\label{navg3}
\end{equation}
This provides an exact expression for the average number of parts.

To proceed, we use a well-known approximation \cite{Nathanson2000,Almkvist2002} for the restricted partition number
\begin{equation}
P_H(X) \approx \frac{1}{(N-1)!} \frac{X^{N-1}}{x_2 \cdots x_N},
\end{equation}
valid for large $X$.  Substituting this approximation into Eq. (\ref{navg3}), removing the floor function and replacing the summations by integrals, we find
\begin{eqnarray}
\langle n_j \rangle &\approx& \frac{1}{X^{N-1}} \int_{1}^{X/x_j} (X - z x_j)^{N-1} dz \nonumber \\
&=& \frac{1}{N x_j} \frac{(X-x_j)^N}{X^{N-1}} \nonumber \\
& \approx & \frac{X}{N x_j} \left[ 1 + \mathcal{O}(X^{-1}) \right].
\end{eqnarray}
A rigorous calculation (including the dependence of the error term on the part set $H$), found by bounding the partition number more precisely, is presented in the Appendix.

\section{Benford's Law}
As described above, power laws, such as Zipf's law, or other ``fat'' or ``long-tailed'' distributions have been studied intensively \cite{Newman05}.  Here we have found the simplest power law for the number of parts of a given size, in the equipartition limit Eq. (\ref{keyeq1}).  Note that the power law is for the average number of parts, as opposed to the probability distribution for the number of each part (which is exponential).  In some sense, this can be seen as the simplest possible distribution, as only one conservation constraint has been imposed on the number of parts.  Most importantly, this simplest power law leads directly to Benford's law \cite{Lemons86}, which we reproduce here for completeness.

Specifically, if we extend the equipartition result Eq. (\ref{keyeq1}) to a system in which we can sample from a continuous set of pieces of size $x$, each occurring with probability proportional to $1/x$, the expected digit distribution (over any interval $10^p \to 10^{p+1}$ in $x$) will be Benford:
\begin{equation}
P_d = \frac{ \int_{ d 10^p}^{(d+1) 10^p} dx /x }{ \int_{10^p}^{10^{p+1}} dx /x} = \frac{ \ln (1+1/d) }{ \ln 10} = \log_{10} \left( 1 + \frac{1}{d} \right).
\end{equation}
We note that other long-tailed distributions may exhibit Benford-like behavior \cite{Pietronero2001}, and thus many of the distributions recently studied \cite{Baek11,Peterson13, Visser13} may also be candidates to describe how Benford-like data sets emerge, but the inverse power law shown here uniquely leads to the exact Benford distribution.

As an example of an almost Benford distribution, we consider the the maximum entropy result Eq. (\ref{navg1}), for continuous $x$.  For this distribution we can perform a similar calculation to find $P_d$, and find that 
\begin{eqnarray}
P_d &=& \frac{\ln \left[ (1-e^{-\beta (d+1) 10^p}) (1 - e^{-\beta d 10^p})^{-1} \right] }{ \ln \left[ (1-e^{-\beta 10^{p+1}}) (1-e^{-\beta 10^p})^{-1} \right] }\nonumber \\
 & \approx & \log_{10} \left(1 + \frac{1}{d} \right) \nonumber \\
 & & + \beta \frac{10^p}{2 \ln 10} \left[ 9 \log_{10} \left(1 + \frac{1}{d}\right) - 1\right]  \ \mbox{for} \ \beta \ll 10^{-p}, \nonumber \\
\end{eqnarray}
so that, in the equipartition limit $\beta \to 0$, we recover the Benford digit distribution.  For large $\beta$, the digit distribution tends to the exponential form $e^{- \beta 10^p d}$.

\section{Beyond Benford: General Power Laws}

We now consider an extension of the maximum entropy principle to allow for arbitrary power-law distributions, along with natural cutoffs.   First, we modify the conservation law of Eq. (\ref{conv_law}) to a more general form
\begin{equation}
X = \sum_{j=1}^N n_j x_j^a.
\end{equation}
This equation can be interpreted geometrically, so that $a=1$ is like the partitioning of a line, $a = 2$ the partitioning of an area, and general $a$ would correspond to more graph-like fractal geometries.  In addition to this generalization, we add a chemical potential $\mu$, corresponding to an average total number of fragments $\mathcal{N} = \langle n_1 + \cdots \rangle$:
\begin{eqnarray}
\mathcal{S} &=& - \sum_{\vec{n}} p(\vec{n}) \ln p(\vec{n})  - \alpha \left( \sum_{\vec{n}} p(\vec{n})- 1\right) \nonumber \\
& & - \beta \left( \sum_{\vec{n}}  p(\vec{n}) \left[ \sum_j n_j x_j^{a} - X  \right] \right) \nonumber \\
& & - \mu \left( \sum_{\vec{n}} p(\vec{n}) \left[ \sum_j n_j  - \mathcal{N} \right] \right).
\end{eqnarray}  

Maximizing this entropy yields the probability distribution
\begin{equation}
p(\vec{n}) = \prod_{j=1}^N (1 - e^{-\beta x_j^{a} - \mu}) e^{-\beta n_j x_j^{a} - n_j \mu}.
\label{prob_gen}
\end{equation}
When we calculate the fragment distribution for this probability, we now find
\begin{equation}
\langle n_j \rangle = \frac{1}{e^{\beta x_j^{a} + \mu} - 1}.
\label{navg_gen}
\end{equation}
This function, a generalized distribution $n(x)$ for fragments of size $x$, can be used to study many empirical data sets with power-law regions.  Specifically, this function has the following properties, as shown in Fig. 2.  First, when $x \ll x_1 \equiv (\mu/\beta)^{1/a}$, the fragment distribution becomes a constant $n(x) \approx 1/(e^{\mu}-1)$, which will be finite for $\mu>0$.  Second, for $x_1 < x < x_2 \equiv \beta^{-1/a}$, $n(x)$ is approximately a power law $n(x) \sim x^{-a}$.  Finally, for $x \gg x_2$, the distribution falls exponentially $n(x) \approx e^{-\beta x^{a}}$.  Thus, this is a normalizable distribution with three characteristic regions generalizing both the exponential and power-law distributions.  

\begin{figure}
\includegraphics[width=3 in]{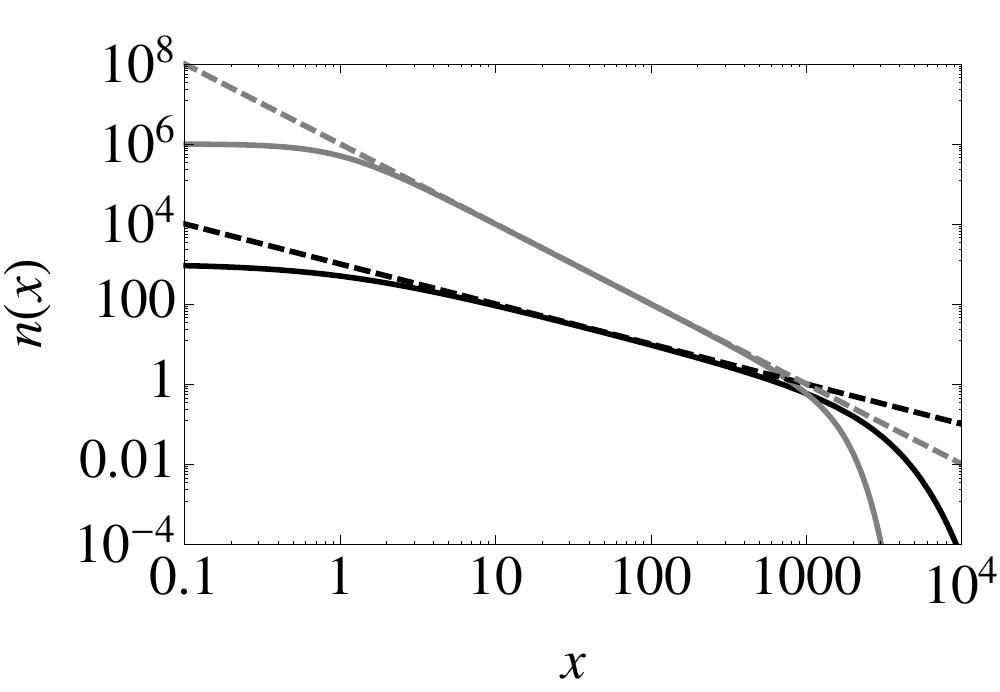}
\label{powerfig}
\caption{Example fragment distributions using the generalized power law Eq. (\ref{navg_gen}), with $x_1 = 1$, $x_2 = 1000$ (see text) and $a = 1$ (black) and $a = 2$ (gray).  Also shown are the corresponding power laws (dashed).  }
\end{figure}

This generalized power-law distribution has many similarities to those derived from maximum entropy subject to alternative constraints \cite{Baek11, Peterson13, Visser13}.  However, instead of directly constructing a probability distribution, we use the probability of obtaining a particular part set [given by Eq. (\ref{prob_gen})] to find the average number of parts [given by Eq. (\ref{navg_gen})].  It is the latter which produces a generalized power law.  This alternative route to generalized power laws may be appropriate for data found by averaging over many realizations.

\section{Conclusion}

In this paper we have explored the question of the distribution of numbers arising from the partitioning of a quantity $X$ into a set of pieces $x_j$.  We have found, using maximum entropy and exact counting, that the average number of parts of size $x_j$ tends to the equipartition result $\langle n_j\rangle = X/ ( N x_j)$ when $X \gg 1$.  This result is intimately related to the statistical mechanics of quantum oscillators and their high-temperature limit.  Extending this result to a continuous set of parts provides an attractive route to Benford's law, an empirical observation regarding the first digits of many real-world data sets.  Finally,  this type of model can be used to generate long-tailed distributions using a small number of parameters, also relevant to many real-world data sets.  Here we consider the open questions regarding the application to Benford's law.

First, the limiting process to go from a discrete set of $N$ parts to a continuous set requires us to specify how both $X$ and $N$ tend to infinity.  The rigorous bounds derived in the Appendix require that $X \gg N^2 x_N$, while careful analysis of the maximum entropy result suggests that a weaker condition of $X \gg N x_N$ may be possible.  Understanding exactly when the equipartition result and Benford's law is truly applicable remains an open question. This is relevant to whether the model presented here is truly applicable to real-world data sets such as the division of large population ($X$) into groups of various sizes ($\{x_j\}$).  Second, the character of the generalized power laws remains obscure.  It would be nice to have a more physical interpretation of the conservation law, and whether it has connection to the other generalized power laws discussed in the literature \cite{Newman05,Baek11, Peterson13, Visser13}.  The variety of these results suggests that there may be many processes underlying these distributions.  This raises a final question, whether the fully random partitioning considered here corresponds to any real-world process.  

Regarding this final point, we have recently analyzed multiple random fragmentation scenarios for a one-dimensional object, similar to those studied in \cite{Frontera95,Yamamoto2012}, and find that the resulting fragments obey Benford's law in the long-time limit \cite{Becker13}.  The convergence of such a process is currently under investigation.  We hope that continued statistical analysis of these problems will help shed light on how power laws and Benford's law emerge in such varied phenomena in nature.

\acknowledgements

This work was supported by Williams College and SJM was supported by Grant DMS1265673 of the National Science Foundation.

\appendix*

\section{Partition Number Bounds}

In this Appendix, we provide bounds on the partition number $P_H(X)$ and the average number of parts $\langle n_j\rangle$.  We begin by observing that the exact restricted partition number can be written as an explicit sum over all possible partitions:
\begin{equation}
\label{eq:fullPHX}
P_{H}(X) = \sum_{n_{N} = 0}^{\lfloor L_N \rfloor}\ \sum_{n_{N-1} = 0}^{\lfloor L_{N-1} \lfloor} \cdots \sum_{n_2 = 0}^{\lfloor L_2 \rfloor}\ \sum_{n_1 = 0}^{\lfloor L_1 \rfloor} \delta\left( X, \sum_{h \in H} n_h x_h\right),
\end{equation}
where the upper limits $\lfloor L_k \rfloor$ denote the maximum number of $n_k$ that can be subtracted from the remainder of $X$, namely
\begin{equation}
\label{eq:lsubk}
L_{k} = \frac{X - \sum_{j = k+1}^{N} n_j x_j}{x_k}.
\end{equation}
For this calculation, we will only consider those sets $H$ such that $x_1 = 1$. This ensures that a partition will exist for every $X$ and allows us to sum over the delta function. We then can disregard the sum over $n_1$, as once the other $n_j$ are determined, it will only have one possible value.  Thus we consider
\begin{equation}
\label{eq:fullPHX}
P_{H}(X) = \sum_{n_{N} = 0}^{\lfloor L_N \rfloor}\ \sum_{n_{N-1} = 0}^{\lfloor L_{N-1} \rfloor} \cdots \sum_{n_2 = 0}^{\lfloor L_2 \rfloor}1.
\end{equation}
We now find upper and lower bounds for $P_H(X)$.

We begin with a lower bound.  Given that our summands $f(n_j)$ are all positive and non-increasing, we have the inequality (A.12 in \cite{AlgorithmsBook})
\begin{equation}
\label{eq:inequal1}
\sum_{n=0}^{\lfloor L \rfloor} f(n) \ge \int_0^{\lfloor L \rfloor+1} f(n) dn  >\int_0^{L} f(n) dn,
\end{equation}
where we have used the fact that
$\lfloor L \rfloor > L-1.$ It follows, then, that
\begin{equation}
\label{eq:intLowPHX}
P_H(X) > \int_{n_{N} = 0}^{L_N}\ \int_{n_{N-1} = 0}^{L_{N-1}} ... \int_{n_2 = 0}^{L_2}\ dn_{2} \cdots dn_{N}.
\end{equation}
It is fairly straightforward to integrate this expression, using a recursion relation [from Eq. (\ref{eq:lsubk})]
\bb
L_{k} = \frac{x_{k+1}}{x_{k}} (L_{k+1} - n_{k+1}), & & k = 2 \to N-1,
\ee
to find
\begin{equation}
\label{eq:lowbound}
P_{H}(X) > \frac{1}{(N-1)!} \frac{X^{N - 1}}{x_2 \cdots x_N}.
\end{equation}

For the upper bound, we again convert our sums into integrals.  Here, however, we use the alternative inequality  (A.12 in \cite{AlgorithmsBook})
\begin{eqnarray}
\label{eq:inequal2}
\sum_{n=0}^{\lfloor L \rfloor} f(n)&\le& \int_{-1}^{\lfloor L \rfloor} f(n) dn \nonumber \\
 &\le& \int_{-1}^{L} f(n) dn \nonumber \\
&=& \int_{0}^{L+1} f(n'-1) dn',
\end{eqnarray}
where $\lfloor L \rfloor \le L$ and we have changed variables $n' = n + 1$.  In terms of these variables, we note that 
\begin{equation}
\label{eq:lsubkPrime}
L_k + 1 = \frac{1}{x_k} \left( X + \sum_{j=k}^N x_j - \sum_{j=k+1}^N n'_j x_j \right).
\end{equation}
We thus use one more inequality 
\begin{equation}
\label{eq:primeInequal}
 L_k + 1 \le L'_k \equiv \frac{1}{x_k} \left( X' - \sum_{j=k+1}^N n'_j x_j \right),
\end{equation}
where
\begin{equation}
\label{eq:xPrime}
X' = X + \sum_{j=2}^N x_j
\end{equation}
and the equality occurs for $k=2$.
Altogether we find
\begin{equation}
\label{eq:int2PHX}
P_{H}(X) < \int_{n'_{N} = 0}^{L'_N}\ \int_{n'_{N-1} = 0}^{L'_{N-1}} \cdots \int_{n'_2 = 0}^{L'_2}\ dn'_{2} \cdots dn'_{N}.
\end{equation}
These integrals can be evaluated as in the lower bound case to yield
\begin{eqnarray}
\label{eq:upperbound}
P_{H}(X) &<& \frac{1}{(N-1)!} \frac{ {X'}^{N-1}}{x_2 \cdots x_N} \nonumber \\
&=& \frac{1}{(N-1)!} \frac{1}{x_2 \cdots x_N} \left( X + \sum_{j=2}^N x_j \right)^{N-1}. \nonumber \\ 
\end{eqnarray}

Having bounded the partition number, we can provide upper and lower bounds for  $\langle n_j \rangle$, using
\begin{equation}
\langle n_j \rangle = \frac{1}{P_{H}(X)} \sum_{\ell = 1}^{\lfloor X/x_j \rfloor} P_{H}(X - \ell x_j).
\label{navg3x}
\end{equation}
To get a lower bound for $\langle n_j \rangle$, we use the upper bound for $P_{H}(X)$ in the denominator and the lower bound for $P_{H}(X)$ in the sum.  Using Eqs. (\ref{eq:lowbound}) and (\ref{eq:upperbound}) in Eq. (\ref{navg3x}), we get
\bb
\label{eq:lownj}
\nonumber \langle n_j \rangle &>& \frac{1}{{X'}^{N-1} } \sum_{\ell = 1}^{\lfloor X / x_j \rfloor} (X - \ell x_j)^{N - 1} \\
&>& \frac{X}{N x_j} \left( 1 - \frac{x_j}{X} - \frac{1}{X} \sum_{k=2}^N x_k \right)^N,
\ee
where we have used the inequalities $(1+x)^{-1} > 1-x$ and Eq. (\ref{eq:inequal1}) for the summation.

To get an upper bound for $\langle n_j \rangle$, we use the lower bound for $P_{H}(X)$ in the denominator and the upper bound for $P_{H}(X)$ in the sum.  Again, using Eqs. (\ref{eq:lowbound}) and (\ref{eq:upperbound})  in Eq. (\ref{navg3x}), we have
\bb
\nonumber \langle n_j \rangle &<& \frac{1}{X^{N-1}} \sum_{\ell = 1}^{\lfloor X / x_j \rfloor}  (X' - \ell x_j)^{N - 1} \\
&<& \frac{X}{N x_j} \left(1 + \frac{x_j}{X} + \frac{1}{X} \sum_{k=2}^N  x_k \right)^N,
\label{eq:highnj}
\ee
where here we have used the inequality of Eq. (\ref{eq:inequal2}) for the summation.
\begin{widetext}
Taking Equations (\ref{eq:lownj}) and (\ref{eq:highnj}) together, we conclude that
\begin{equation}
\label{eq:actualbounds}
\frac{X}{N x_j} \left(1 - \frac{x_j}{X} - \frac{1}{X} \sum_{k=2}^N x_k \right)^N < \langle n_j \rangle <  \frac{X}{N x_j} \left( 1 + \frac{x_j}{X} + \frac{1}{X} \sum_{k=2}^N x_k  \right)^N.
\end{equation}
In the large $X$ limit, we Taylor expand each side of Eq. (\ref{eq:actualbounds}) to find
\begin{equation}
\label{eq:approxbounds}
\nonumber \frac{X}{N x_j}\left( 1 - \frac{N x_j}{X} - \frac{N}{X} \sum_{k=2}^N x_k + \cdots \right) < \langle n_j \rangle <  \frac{X}{N x_j} \left(1 + \frac{N x_j}{X} + \frac{N}{X} \sum_{k=2}^N x_k  + \cdots \right).
\end{equation}
We conclude that
\begin{equation}
\label{eq:boundLemons}
\langle n_j \rangle = \frac{X}{N x_j} \left[ 1 + \mathcal{O}(X^{-1}) \right],
\end{equation}
in agreement with the calculations presented in the text.
\end{widetext}

\bibliography{benford}

\end{document}